\documentclass[conference]{IEEEtran}
\IEEEoverridecommandlockouts
\usepackage{cite}
\usepackage{amsmath,amssymb,amsfonts}
\usepackage{algorithmic}
\usepackage{graphicx}
\usepackage{textcomp}
\usepackage[table,xcdraw]{xcolor}
\usepackage{balance}
\usepackage{pifont}
\newcommand{\cmark}{\ding{51}}%
\newcommand{\xmark}{\ding{55}}%
\def\BibTeX{{\rm B\kern-.05em{\sc i\kern-.025em b}\kern-.08em
    T\kern-.1667em\lower.7ex\hbox{E}\kern-.125emX}}
\begin{document}

\title{UCIL: An Unsupervised Class Incremental Learning Approach for Sound Event Detection}

\author{\IEEEauthorblockN{Yang Xiao and Rohan Kumar Das}
\IEEEauthorblockA{\textit{Fortemedia Singapore, Singapore} \\
Email: \{xiaoyang, rohankd\}@fortemedia.com}}
\maketitle

\begin{abstract}
This work explores class-incremental learning (CIL)  for sound event detection (SED), advancing adaptability towards real-world scenarios. CIL's success in domains like computer vision inspired our SED-tailored method, addressing the unique challenges of diverse and complex audio environments. Our approach employs an independent unsupervised learning framework with a distillation loss function to integrate new sound classes while preserving the SED model consistency across incremental tasks. We further enhance this framework with a sample selection strategy for unlabeled data and a balanced exemplar update mechanism, ensuring varied and illustrative sound representations. Evaluating various continual learning methods on the DCASE 2023 Task 4 dataset, our research offers insights into each method's applicability for real-world SED systems that can have newly added sound classes. The findings also delineate future directions of CIL in dynamic audio settings.
\end{abstract}


\section{Introduction}
\label{sec:intro}
Sound event detection (SED)~\cite{sed,sed2} identifies and classifies various sounds within an audio stream. It supports applications like urban security and wildlife monitoring by detecting specific noises~\cite{home,smarthome,safety}. SED parses continuous audio, segments it, and labels each event with predefined sound classes. Deep learning has greatly improved SED, making it effective in controlled environments. These systems generally operate under a closed set framework with a fixed dataset comprising predetermined sound classes but real-world scenarios require incrementally recognizing new sound events~\cite{wild}.

In response to the evolving nature of sound events in practical applications, different strategies have been explored to incorporate new sound classes into existing SED models. The most common approach is fine-tuning~\cite{ft1,ft2,uaed}, where a pretrained model is retrained using a smaller dataset of new classes. While fine-tuning helps integrate new knowledge~\cite{ft5,ft3,ft4}, it can negatively impact the model's performance in previously learned classes due to catastrophic forgetting.
This challenge arises because the model struggles to learn new classes without losing accuracy on the ones it already knows.
The practical issue is not just about introducing new classes; it's about enabling the model to expand its abilities over time while still holding onto the knowledge it initially acquired.


The above context highlights the need for continual learning~\cite{cl,cl1}, a paradigm that enables models to continuously acquire new knowledge while preserving existing information. This approach addresses catastrophic forgetting, ensuring models can adapt and evolve effectively over time. In continual learning, class-incremental learning (CIL)~\cite{cil1} integrates new classes into a model’s learning architecture sequentially. CIL is pivotal in retaining previously learned knowledge, proving essential for applications requiring ongoing model evolution. It has been effective in areas such as computer vision (CV)~\cite{clcv,clav} and natural language processing\cite{clnlp}, significantly contributing to their advancements. 

For CIL's extension into the audio~\cite{claudio} and speech domain~\cite{cl2}, covering tasks like audio classification (AC)~\cite{cl4,cdoa,cl5} and keyword spotting\cite{cl3,clkws2}, has proven the method's adaptability and utility in managing dynamic datasets. Inspired with~\cite{lucir,icarl}, Mulimani et al.~\cite{cl4,cl5} introduce one independent learning method with two distillation losses for the CIL of multi-label audio classification. However, their approach is for audio classification with clip-wise labels and does not compare with other continual learning methods. Although strongly labeled SED has been explored much in recent years under the DCASE challenge Task 4~\cite{desed}, the CIL application in SED with strong labels to estimate the precise onset and offset is relatively recent. While Koh et al.~\cite{clsed1} have made strides in adapting models for SED through transfer learning, it does not cover the scope for incremental learning to develop real-world strong labeled SED systems. 

This work introduces a novel CIL method for SED that aims to incorporate new sound categories and maintain accurate detection of sound events. Drawing inspiration from techniques used in~\cite{lucir,cl4}, we implemented {\it an unsupervised class incremental learning (UCIL) framework} to efficiently train the classifier on different tasks. We integrate a distillation loss function that helps in retaining the classifier's knowledge of previously learned sound events while learning new ones, minimizing feature differences between model iterations. Because the strong label is limited, our strategy also includes a sample selection process to {\it leverage unlabeled data} for improving the consistency of the SED model. 

Further, we propose a technique to {\it preserve previous examples for future rehearsal} by considering the duration and frequency of events within each sound category ensuring a balance. The proposed UCIL is evaluated on the DCASE 2023 Task 4A dataset~\cite{desed} with comparison to some CIL methods. 
To the best of our knowledge, this is the first work to apply {\it rehearsal-based CIL techniques to the strongly labeled SED}, thereby opening the door for future research directions in real-world scenarios for SED applications.
\section{Class Incremental Learning}

\subsection{Class Incremental Learning for SED}

We introduce a CIL approach for SED, where the model learns to recognize a variety of sounds expanded through a series of tasks $t_0, t_1, \ldots, t_T$ each representing a different set of sounds to be identified. This learning journey unfolds over $1 + T$ time steps, starting with an initial task $t_0$ and moving through $T$ subsequent tasks, each introducing new sounds for detection. Every task $t_i$ has its own dataset $D_i$, consisting of sound features $x_j^i$ and their corresponding labels $y_j^i$ that are represented as multi-hot vectors. The total number of sound classes $|C|$ evolves to include both existing classes learned in previous tasks ($|C_{\text{ext}}|$) and the new classes introduced in the current task ($|C_{\text{cur}}|$). Our model, designated as $M_i$ for each task $t_i$, comprises two main components: a feature extractor $\theta_i$ and a classifier $\phi_i$. The feature extractor analyzes the sound data to identify distinguishing characteristics, while the classifier uses these features to predict the types of sounds present. 

\subsection{Proposed UCIL Method}

The proposed UCIL method comprises four integral components, which we discuss in the following. 

\subsubsection{Independent Learning to Update Model by New Data}
We start by training a learning model on an initial task, which we refer to as $t_0$. This initial task uses a dataset named $D_0$. We employ binary cross-entropy loss ($\mathcal{L}_{cls}$) that computes the difference between the model predictions and the ground truth. It operates by applying a function, the sigmoid function $\sigma$, to the model's logits $o_k$ in class $k$ as:
\begin{equation}
\begin{aligned}
    \mathcal{L}_{cls} =  -\frac{1}{|C|} & \sum_{k=1}^{|C|}  y_k^i \cdot \log(\sigma(o_k)) \\ &  +  (1 - y_k^i) \cdot \log(1 - \sigma(o_k)),
\end{aligned}
\vspace{-1mm}
\end{equation}

After the initial task $t_0$, we create an updated version of our model denoted as $M_i$ by new sounds in subsequent tasks $t_i$. We add the new classification heads of the linear layer, building on the previous model $\hat{M}_{i-1}$ to accommodate the new sound classes $|C_{\text{cur}}|$ of the current task. 
The predictions made by the model $M_i$ are now a mix of existing and new sound classes, which we refer to as $o_{ext}$ and $o_{cur}$, respectively. The updated model, with its new feature extractor $\hat{\theta}_i$ and classifier $\hat{\phi}_i$, is then trained on the latest data set, $\hat{D}_i$. To prevent data imbalance issues of existing and new classes, we use the independent learning (IndL)~\cite{cl5} method, which allows the model to learn the new sound predictions independently from the existing ones. Specifically, we only compute the $\mathcal{L}_{cls}$ by only the new predictions. This way, the existing predictions are maintained separately.

\subsubsection{Knowledge Distillation from Existing to New}
\label{sec222}
We use two special distillation losses to maintain the learned knowledge of our model~\cite{lucir}. The first one is prediction distillation loss $\mathcal{L}_{dis}^P$, which reduces differences in the model's initial predictions for known sound classes. We use mean squared error (MSE) loss to effectively handle the precise temporal alignment required in strongly labeled SED. The second one is the feature distillation loss $\mathcal{L}_{dis}^F$, which minimizes the gap between the current and existing model's features by comparing their alignment with cosine similarity. The current model $M_i$'s feature is the output of the feature extractor $\theta_i$ given as $v=\theta_i(x)$. In simpler terms, these techniques help our updated model $M_i$ act like its existing version, $\hat{M}_{i-1}$, especially for sounds it already learned.
\vspace{-2mm}
\begin{equation}
\begin{aligned}
& \mathcal{L}_{dis}^P = \lambda MSE(\hat{M}_{i-1}(x) || o_{ext}) \\
& \mathcal{L}_{dis}^F = 1 - \cos(\text{norm}(v), \text{norm}(\bar{v})) 
\end{aligned}
\vspace{-2mm}
\end{equation}
where $\text{norm}$ and $\cos$ denote the L2-normalization and cosine similarity, respectively. In detail, we compute the scalar product with normalized vectors to get the cosine similarity. The weight $\lambda = \Omega \sqrt{\frac{|C|}{|C_{cur}|}}$ is set adaptively, $\Omega$ is a fixed constant, and $\lambda$ increases as the number of incremental phases increases, to retain the knowledge of the increasing number of existing classes in the model. For the existing sounds, we measure the output predictions from $M_i$ and match them against $\hat{M}_{i-1}$. 


\begin{figure*}[t!]
  \centering
  \vspace{-10mm}
  \includegraphics[width=0.9\linewidth]{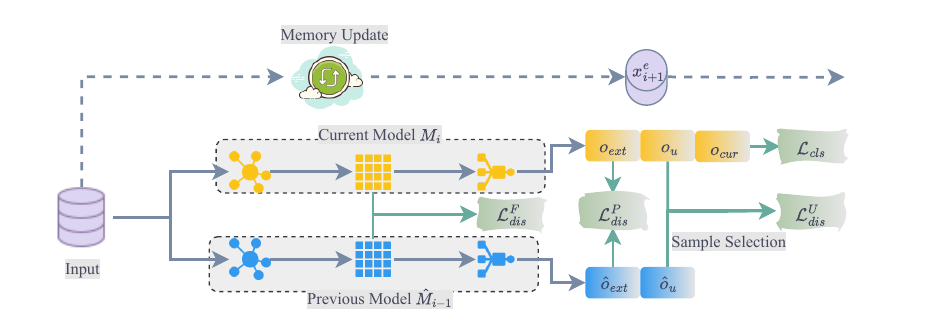}
  \vspace{-6mm}
  \caption{Block diagram of the proposed UCIL approach for task $t_i$. All training data for task $t_i$ include the new data $D_i$, the rehearsal data $x^e_i$, and the unlabeled data $x^u_i$ as the input. $o_{ext}$, $o_{cur}$, and $o_{u}$ present the prediction of the model $M_i$ for the classes of the existing task, classes of the current learning task, and the unlabeled data.}
  \label{fig:workflow}
  \vspace{-6mm}
\end{figure*}

\subsubsection{Unsupervised Learning with Sample Selection}

Given the limited availability of strongly labeled data, researchers have proposed various SED models that incorporate unlabeled data through semi-supervised learning. When unlabeled data influences SED model behavior, it becomes crucial to leverage these predictions to maintain model consistency. 
Therefore, we introduce unlabeled data $x^u_i$ into the dataset $D_i$ of the task $t_i$. However, not all data is helpful—some can confuse the model. Hence, we come up with a novel way to pick the helpful samples from unlabeled data. We look for samples where the model predictions differ greatly due to forgetting, assuming these samples will teach the model effectively. Specifically, we measure the discrepancy between two logit distributions with the L1 metric as:
\vspace{-2mm}
\begin{equation}
d = |\hat{M}_{i-1}(x^u_i) - M_{i}(x^u_i)|
\vspace{-2mm}
\end{equation}

We consider the first half of samples based on the discrepancy $d$ from our unlabeled data for each batch of training task $t_i$. By selecting these samples by discrepancy, we assist the model to retain previously learned knowledge and generalize from the past knowledge to the new situations. We adopt MSE loss given in Section~\ref{sec222} to use as $\mathcal{L}_{dis}^U$, focusing on unlabeled data. Our goal is to reduce the gap between the new and existing models for the reaction of unlabeled data.


\subsubsection{Balanced Memory Update Method}
To further maintain the previous knowledge,
especially in applications with complex classes such as SED, we preserve some of the data from existing classes in advance as the rehearsal data. We mix this data $x^e_i$ with the incoming data of task $t_i$. The balanced sampling function is an algorithm tailored for preparing rehearsal data in the context of SED using strongly and weakly labeled data. It starts by computing the duration of sound events, considering the difference between their onset and offset times. Then, it aggregates these durations for each event, providing a measure of total exposure for each sound in the strongly labeled dataset. For weakly labeled data, the duration is 10 seconds for each event label. 

The method makes sure that there is a fair mix of sounds for training the model. It figures out how many examples of each sound are needed based on the seen classes. It then picks a balanced set of sound examples from both types of data to avoid any bias. The outcome is two rehearsal datasets, one from strongly labeled and the other from weakly labeled data. This way, all seen sounds get a fair chance to replay during training, leading to a more reliable and fair SED model. 

The approach for independent unsupervised learning as illustrated in Fig. \ref{fig:workflow}. The total loss function \(\mathcal{L}_{tot}\) of the updated model is given as: 
\vspace{-2mm}
\begin{equation}
    \mathcal{L}_{tot} = \mathcal{L}_{cls} + \mathcal{L}_{dis}^P + \mathcal{L}_{dis}^F + \mathcal{L}_{dis}^U
\vspace{-2mm}
\end{equation}
In summary, the model is first trained on one task and then updated with new classifier units for enhanced classification. Independent learning separates the learning of new and existing sound classes. To retain knowledge, we use prediction and feature distillation losses, ensuring consistency across classes. We also strategically select samples from unlabeled data to improve generalization. Finally, our balanced memory update method prepares rehearsal data from both strongly and weakly labeled datasets, ensuring fair representation.
\section{Experiment Setting}
\subsection{Dataset, Task Setting and Performance Metric}
In this study, we use the DCASE 2023 Task 4A dataset~\cite{desed}, a comprehensive collection of 10-second audio clips that are categorized into 10 sound events. It is comprised of 1,578 real recordings with weak annotations, 14,412 real recordings without labels, and 10,000 synthetic recordings strongly labeled~\cite{impact}. All clips were resampled to 16 kHz mono and segmented using a 2048-sample window and 256-sample hop length for log-mel spectrogram generation.

We devised two incremental task settings to assess our proposed method. Firstly, the two-task setting randomly divides the 10 target sound classes in the dataset into two groups, each containing 5 classes. This approach is designed to test how well our method can adapt from one group of sounds to another, a process known as source-target domain adaptation in SED. Secondly, the four-task setting organizes the 10 target classes into 4 groups randomly where two groups contain two sound events and the rest two contain three sound events. We shuffle the groups as the four tasks to investigate the effect of the number of tasks on the model's ability to distinguish between different sounds, providing insights into the complexities of learning diverse acoustic events.

We used the threshold independent polyphonic sound event detection scores (PSDS)~\cite{psds,tpsds} metric following the DCASE 2023 Task 4A protocol across two scenarios for system evaluation. PSDS1 evaluates the system's ability for precise timing in detecting sound events, crucial for real-time applications, whereas PSDS2 focuses on the system's capability to distinguish between similar sound classes, addressing class confusion in SED. The final performance for the two settings is evaluated after the last task is learned.


\subsection{Implementation Details and Reference Baselines}
We used a batch size of 48 and the Adam~\cite{adam} optimizer with an initial 0.001 learning rate, gradually increasing over the first 50 of 200 total epochs following the DCASE 2023 Task 4A setup. We adopted a mean teacher model strategy~\cite{meanteacher} to enhance training stability, applying an exponential moving average with a factor of 0.999. We consider a convolutional recurrent neural network (CRNN)~\cite{crnn} model that serves as the DCASE baseline with about 1.2M parameters for all the studies in this work. We implemented early stopping at 50 epochs to prevent overfitting, and set the weight $\Omega$ for the prediction distillation loss ($\mathcal{L}_{dis}^P$) at 2, ensuring a balance between retaining existing knowledge and learning new information. We built following 5 reference baselines:

\begin{table}[t]
\vspace{-2mm}
\centering
\caption{Performance comparison of our proposed UCIL method with other CIL methods using the two-task settings. }
\vspace{-2mm}
\label{tab:two--table}
\resizebox{0.8\linewidth}{!}{%
\begin{tabular}{cccc}
\hline
\textbf{Method}   & \textbf{Rehearsal Size}   & \textbf{PSDS1} & \textbf{PSDS2} \\ \hline\
Finetune (lower bound) & \textbackslash{} &   0    &    0.086   \\ \hline
Joint (upper bound)   & \textbackslash{} &   0.354    &    0.565   \\ \hline
EWC      & {[}4000, 400{]}  &   0.255    &   0.443    \\ 
LwF      & {[}4000, 400{]}  &   0.247    &   0.459   \\ 
NR       & {[}4000, 400{]}  &   0.216   &    0.416   \\ \hline
UCIL    & {[}4000, 400{]}  &   \cellcolor[HTML]{C4D5EB}{\bf 0.260}    &   \cellcolor[HTML]{C4D5EB}{\bf 0.466}    \\ 
UCIL    & {[}2000, 200{]}  &   0.237   &    0.434   \\ 
UCIL    & {[}1000, 100{]}  &   0.217    &   0.417    \\ \hline
\end{tabular}%
}
\vspace{-5mm}
\end{table}


\begin{itemize}
\item \textbf{Fine-tune Training:} It acts as our lower-bound baseline, where the CRNN model is simply adjusted for new tasks without any strategies to prevent forgetting existing classes, demonstrating minimum expected performance.
\item \textbf{NR (Naive Rehearsal)~\cite{nr}:} It randomly selects samples from previous tasks for retraining, aiming to maintain memory through repetition.
\item \textbf{EWC (Elastic Weight Consolidation)~\cite{ewc}:} This introduces a penalty on changing important parameters for past tasks, determined by the Fisher information matrix, to balance new learning and memory retention.
\item \textbf{LwF (Learning Without Forgetting)~\cite{lwf}:} It updates the model for new tasks while ensuring existing class performance is maintained by training on new classes and reinforcing consistent predictions for existing ones.

\item \textbf{Joint Training:} Our upper-bound baseline trains the CRNN model with the entire dataset, showing the maximum potential performance without incremental task setting. In incremental task setting, the performance can never surpass the joint training as it already knows the whole target classes the first time. 
\end{itemize}
\section{Results and Analysis}
\subsection{Results for Two-task and Four-Task Settings}
We first conduct the studies under the two-task setting on the DCASE 2023 Task 4A dataset, whose results are reported in Table~\ref{tab:two--table}. Rehearsal size indicates the maximum clip numbers of [Strongly Labeled, Weakly Labeled] as the rehearsal data in each task. We consider random sampling to add the rehearsal data in EWC and LwF methods for a fair comparison. We focus on demonstrating how CIL minimizes the performance gap due to catastrophic forgetting.
It observed that the lower-bound finetune approach faces challenges in adapting to new tasks without retention strategies as expected. The PSDS score is observed to be zero which indicates the model forgot most of the previous knowledge and similar observation was also reported in~\cite{cl5}. Analyzing the impact of different CIL strategies between scenarios, it is evident that each method's ability to handle the nuances of prompt detection (PSDS1) versus class confusion (PSDS2) varies. Our UCIL method, particularly with the highest rehearsal size, shows notable performance, closer to the upper bound in both metrics. This highlights the competence of UCIL in rapidly reacting to new sounds while effectively managing class distinctions. The performance of UCIL method across different rehearsal sizes also provides insights into its scalability. While a decrease in rehearsal size leads to a slight reduction in PSDS, UCIL maintains competitive performance. This is particularly significant on comparing UCIL to other CIL approaches such as EWC, LwF, and NR, where UCIL consistently outperforms or matches their scores, specifically for PSDS2.

\begin{table}[t]
\centering
\caption{Performance comparison of our proposed UCIL method with other CIL methods using the four-task settings. Rehearsal size is varied similar to two-task setting.}
\vspace{-2mm}
\label{tab:four--table}
\resizebox{0.9\linewidth}{!}{%
\begin{tabular}{cccc}
\hline
\textbf{Method}   & \textbf{Rehearsal Size}   & \textbf{PSDS1} & \textbf{PSDS2} \\ \hline
Finetune (lower bound) & \textbackslash{} &   0    &    0.054   \\ \hline
Joint (upper bound)   & \textbackslash{} &   0.354    &    0.565   \\ \hline
EWC      & {[}2000, 200{]}  &   0.137    &   0.251    \\ 
LwF      & {[}2000, 200{]}  &    0.107   &    0.172   \\ 
NR       & {[}2000, 200{]}  &   0.100    &    0.152   \\ \hline
UCIL    & {[}2000, 200{]}  &   \cellcolor[HTML]{C4D5EB}{\bf 0.198}   &    \cellcolor[HTML]{C4D5EB}{\bf 0.366}   \\ 
UCIL    & {[}1000, 200{]}  &   0.184    &   0.332   \\ 
UCIL    & {[}1000, 100{]}  &   0.165    &   0.319    \\ \hline
\end{tabular}%
}
\vspace{-5mm}
\end{table}


We then conduct the studies for four-task setting that are reported in Table~\ref{tab:four--table}.
It is observed that UCIL demonstrates a remarkable performance, especially notable in the PSDS2 scores, where its ability to minimize class confusion gets highlighted. With a rehearsal size of [2000, 200], UCIL's PSDS2 score doubles that of NR and significantly outperforms other CIL methods like EWC and LwF. Moreover, the comparison between UCIL's performance with different rehearsal sizes provides valuable insights into its efficiency. Despite a reduction in the amount of strongly labeled data, UCIL maintains commendable performance, showcasing its effective use of data under a higher task number based CIL studies.

\begin{table}[t]
\centering
\caption{Ablation studies for UCIL method with different settings. All experiments are with [1000, 200] rehearsal data size.}
\vspace{-2mm}
\label{tab:ab-table}
\resizebox{0.8\linewidth}{!}{%
\begin{tabular}{ccccc}
\hline
 \textbf{FD} & \textbf{UL} & \textbf{MU} & \textbf{PSDS1} & \textbf{PSDS2} \\ \hline
 \cmark           &   \cmark          & \cmark           &        \cellcolor[HTML]{C4D5EB}{\bf 0.184}        &      \cellcolor[HTML]{C4D5EB}{\bf 0.332}          \\ 
 \cmark           & \cmark           &      \xmark      &        0.081        &      0.112          \\ 
 \cmark           &     \xmark       & \cmark          &         0.073        &     0.143           \\ 
     \xmark      & \cmark           & \cmark           &        0.110        &        0.193        \\ 

 \cmark           &  w/o Selection         & \cmark           &        0.171        &      0.284          \\ 
 \cmark           & \cmark           & w/o Balance           &        0.135        &      0.217         \\ \hline

\end{tabular}

}
\vspace{-6mm}
\end{table}

\subsection{Ablation Study}

Table~\ref{tab:ab-table} shows an ablation study of our UCIL method, examining the significance of each component under the four-task setting. Each experiment is conducted with a fixed rehearsal size of [1000, 200] for strongly and weakly labeled data, providing a controlled environment to isolate the impact of each module: feature distillation (FD), unsupervised learning (UL), and memory update (MU). We find that the best performance is achieved when all components are active (FD, UL, MU).
Removing MU or even leads to a notable drop in scores, highlighting MU's critical role in effectively integrating new information while retaining previous knowledge. Similarly, excluding UL also decreases performance, underscoring UL's importance in leveraging unlabeled data for broader learning. Excluding FD while keeping UL and MU active shows a lesser but still notable decline in performance, highlighting FD's contribution to maintaining model consistency and effectiveness across varied sound events. Additionally, we observe that removing the (w/o) selection process for unlabeled data results in a moderate drop in performance, indicating the importance of selection. The model struggles more when removing (w/o) balance sampling, suggesting that balanced rehearsal data maintains performance across all sound classes.

\section{Conclusion}
In this work, we proposed a novel method for CIL referred to as UCIL to advance SED research. Incorporating strategies such as independent learning, knowledge distillation and unsupervised sample selection, our method stands out for its ability to maintain model consistency and leverage unlabeled data effectively. The balanced update mechanism further ensures equitable representation of sound classes. The studies on the DCASE 2023 Task 4 dataset, not only showcase the practical applicability of UCIL, but also set a stage for future research in dynamic audio environments.

\clearpage

\balance
\bibliographystyle{IEEEtran}
\bibliography{refs}

\end{document}